\documentclass[final]{svjour3_compact_authors}
\usepackage{graphicx}
\usepackage{rotating}
\usepackage{amssymb}
\usepackage{mathptmx}
\usepackage[numbers]{natbib}
\usepackage{subcaption}
\usepackage{hyperref}
\usepackage{floatrow}
\usepackage{sidecap}
\usepackage{siunitx}

\sidecaptionvpos{figure}{c}
\makeatletter
\journalname{Journal of Low Temperature Physics}

\bibpunct{}{}{,}{s}{}{,}
\begin{document}

\newcommand{\hdblarrow}{H\makebox[0.9ex][l]{$\downdownarrows$}-}
\newcommand*\chem[1]{\ensuremath{\mathrm{#1}}}

\title{Optical Characterization of the SPT-3G Focal Plane}
\def\Cardiff{a}
\def\KIPAC{b}
\def\Stanford{c}
\def\SLAC{d}
\def\FNAL{e}
\def\KICPChicago{f}
\def\NIST{g}
\def\Berkeley{h}
\def\ANLHEP{i}
\def\AAUChicago{j}
\def\EFIChicago{k}
\def\PhysicsUChicago{l}
\def\McGill{m}
\def\ANLMSD{n}
\def\CIFAR{o}
\def\CASA{p}
\def\CaseWestern{q}
\def\Colorado{r}
\def\illast{s}
\def\illphy{t}
\def\LBNL{u}
\def\UChicago{v}
\def\Dunlap{w}
\def\threespeed{x}
\def\CfA{y}
\def\Toronto{z}
\def\UCLA{aa}
\def\supit#1{\raisebox{0.8ex}{\small\it #1}\hspace{0.05em}}

\author{
   Z.~Pan\protect\supit{\KICPChicago,\PhysicsUChicago} \and
   P.~A.~R.~Ade\protect\supit{\Cardiff} \and
   Z.~Ahmed\protect\supit{\KIPAC,\Stanford,\SLAC} \and
   A.~J.~Anderson\protect\supit{\FNAL,\KICPChicago} \and
   J.~E.~Austermann\protect\supit{\NIST} \and
   J.~S.~Avva\protect\supit{\Berkeley} \and
   R.~Basu Thakur\protect\supit{\KICPChicago} \and
   A.~N.~Bender\protect\supit{\ANLHEP,\KICPChicago} \and
   B.~A.~Benson\protect\supit{\FNAL,\KICPChicago,\AAUChicago} \and
   J.~E.~Carlstrom\protect\supit{\KICPChicago,\EFIChicago,\PhysicsUChicago,\ANLHEP,\AAUChicago} \and
   F.~W.~Carter\protect\supit{\ANLHEP,\KICPChicago} \and
   T.~Cecil\protect\supit{\ANLHEP} \and
   C.~L.~Chang\protect\supit{\ANLHEP,\KICPChicago,\AAUChicago} \and
   J.~F.~Cliche\protect\supit{\McGill} \and
   A.~Cukierman\protect\supit{\Berkeley} \and
   E.~V.~Denison\protect\supit{\NIST} \and
   T.~de~Haan\protect\supit{\Berkeley} \and
   J.~Ding\protect\supit{\ANLMSD} \and
   M.~A.~Dobbs\protect\supit{\McGill,\CIFAR} \and
   D.~Dutcher\protect\supit{\KICPChicago,\PhysicsUChicago} \and
   W.~Everett\protect\supit{\CASA} \and
   A.~Foster\protect\supit{\CaseWestern} \and
   R.~N.~Gannon\protect\supit{\ANLMSD} \and
   A.~Gilbert\protect\supit{\McGill} \and
   J.~C.~Groh\protect\supit{\Berkeley} \and
   N.~W.~Halverson\protect\supit{\CASA,\Colorado} \and
   A.~H.~Harke-Hosemann\protect\supit{\illast,\ANLHEP} \and
   N.~L.~Harrington\protect\supit{\Berkeley} \and
   J.~W.~Henning\protect\supit{\KICPChicago} \and
   G.~C.~Hilton\protect\supit{\NIST} \and
   W.~L.~Holzapfel\protect\supit{\Berkeley} \and
   N.~Huang\protect\supit{\Berkeley} \and
   K.~D.~Irwin\protect\supit{\KIPAC,\Stanford,\SLAC} \and
   O.~B.~Jeong\protect\supit{\Berkeley} \and
   M.~Jonas\protect\supit{\FNAL} \and
   T.~Khaire\protect\supit{\ANLMSD} \and
   A.~M.~Kofman\protect\supit{\illphy,\illast} \and
   M.~Korman\protect\supit{\CaseWestern} \and
   D.~Kubik\protect\supit{\FNAL} \and
   S.~Kuhlmann\protect\supit{\ANLHEP} \and
   C.~L.~Kuo\protect\supit{\KIPAC,\Stanford,\SLAC} \and
   A.~T.~Lee\protect\supit{\Berkeley,\LBNL} \and
   A.~E.~Lowitz\protect\supit{\KICPChicago} \and
   S.~S.~Meyer\protect\supit{\KICPChicago,\EFIChicago,\PhysicsUChicago,\AAUChicago} \and
   D.~Michalik\protect\supit{\UChicago} \and
   J.~Montgomery\protect\supit{\McGill} \and
   A.~Nadolski\protect\supit{\illast} \and
   T.~Natoli\protect\supit{\Dunlap} \and
   H.~Nguyen\protect\supit{\FNAL} \and
   G.~I.~Noble\protect\supit{\McGill} \and
   V.~Novosad\protect\supit{\ANLMSD} \and
   S.~Padin\protect\supit{\KICPChicago} \and
   J.~Pearson\protect\supit{\ANLMSD} \and
   C.~M.~Posada\protect\supit{\ANLMSD} \and
   A.~Rahlin\protect\supit{\FNAL,\KICPChicago} \and
   J.~E.~Ruhl\protect\supit{\CaseWestern} \and
   L.~J.~Saunders\protect\supit{\ANLHEP,\KICPChicago} \and
   J.~T.~Sayre\protect\supit{\CASA} \and
   I.~Shirley\protect\supit{\Berkeley} \and
   E.~Shirokoff\protect\supit{\KICPChicago,\AAUChicago} \and
   G.~Smecher\protect\supit{\threespeed} \and
   J.~A.~Sobrin\protect\supit{\KICPChicago,\PhysicsUChicago} \and
   A.~A.~Stark\protect\supit{\CfA} \and
   K.~T.~Story\protect\supit{\KIPAC,\Stanford} \and
   A.~Suzuki\protect\supit{\Berkeley,\LBNL} \and
   Q.~Y.~Tang\protect\supit{\KICPChicago,\AAUChicago} \and
   K.~L.~Thompson\protect\supit{\KIPAC,\Stanford,\SLAC} \and
   C.~Tucker\protect\supit{\Cardiff} \and
   L.~R.~Vale\protect\supit{\NIST} \and
   K.~Vanderlinde\protect\supit{\Dunlap,\Toronto} \and
   J.~D.~Vieira\protect\supit{\illast,\illphy} \and
   G.~Wang\protect\supit{\ANLHEP} \and
   N.~Whitehorn\protect\supit{\UCLA,\Berkeley} \and
   V.~Yefremenko\protect\supit{\ANLHEP} \and
   K.~W.~Yoon\protect\supit{\KIPAC,\Stanford,\SLAC} \and
   M.~R.~Young\protect\supit{\Toronto}
}

\institute{
   \protect\supit{\Cardiff}School of Physics and Astronomy, Cardiff Univ., Cardiff CF24 3YB, United Kingdom \and
   \protect\supit{\KIPAC}Kavli Institute for Particle Astrophysics and Cosmology, Stanford Univ., 452 Lomita Mall, Stanford, CA 94305 \and
   \protect\supit{\Stanford}Dept. of Physics, Stanford Univ., 382 Via Pueblo Mall, Stanford, CA 94305 \and
   \protect\supit{\SLAC}SLAC National Accelerator Laboratory, 2575 Sand Hill Rd., Menlo Park, CA 94025 \and
   \protect\supit{\FNAL}Fermi National Accelerator Laboratory, MS209, P.O. Box 500, Batavia, IL 60510-0500 \and
   \protect\supit{\KICPChicago}Kavli Institute for Cosmological Physics, Univ. of Chicago, 5640 S. Ellis Ave., Chicago, IL 60637 \and
   \protect\supit{\NIST}National Institute of Standards and Technology, 325 Broadway, Boulder, CO 80305 \and
   \protect\supit{\Berkeley}Dept. of Physics, Univ. of California, Berkeley, CA 94720 \and
   \protect\supit{\ANLHEP}Argonne National Laboratory, High-Energy Physics Division, 9700 S. Cass Ave., Argonne, IL 60439 \and
   \protect\supit{\AAUChicago}Dept. of Astronomy and Astrophysics, Univ. of Chicago, 5640 S. Ellis Ave., Chicago, IL 60637 \and
   \protect\supit{\EFIChicago}Enrico Fermi Institute, Univ. of Chicago, 5640 S. Ellis Ave., Chicago, IL 60637 \and
   \protect\supit{\PhysicsUChicago}Dept. of Physics, Univ. of Chicago, 5640 S. Ellis Ave., Chicago, IL 60637 \and
   \protect\supit{\McGill}Dept. of Physics, McGill Univ., 3600 Rue University, Montreal, Quebec H3A 2T8, Canada \and
   \protect\supit{\ANLMSD}Argonne National Laboratory, Material Science Division, 9700 S. Cass Ave., Argonne, IL 60439 \and
   \protect\supit{\CIFAR}Canadian Institute for Advanced Research, CIFAR Program in Cosmology and Gravity, Toronto, ON, M5G 1Z8, Canada \and
   \protect\supit{\CASA}CASA, Dept. of Astrophysical and Planetary Sciences, Univ. of Colorado, Boulder, CO 80309 \and
   \protect\supit{\CaseWestern}Physics Dept., Case Western Reserve Univ., Cleveland, OH 44106 \and
   \protect\supit{\Colorado}Dept. of Physics, Univ. of Colorado, Boulder, CO 80309 \and
   \protect\supit{\illast}Astronomy Dept., Univ. of Illinois, 1002 W. Green St., Urbana, IL 61801 \and
   \protect\supit{\illphy}Dept. of Physics, Univ. of Illinois, 1110 W. Green St., Urbana, IL 61801 \and
   \protect\supit{\LBNL}Physics Division, Lawrence Berkeley National Laboratory, Berkeley, CA 94720 \and
   \protect\supit{\UChicago}Univ. of Chicago, 5640 S. Ellis Ave., Chicago, IL 60637 \and
   \protect\supit{\Dunlap}Dunlap Institute for Astronomy and Astrophysics, Univ. of Toronto, 50 St George St, Toronto, ON, M5S 3H4, Canada \and
   \protect\supit{\threespeed}Three-Speed Logic, Inc., Vancouver, B.C., V6A 2J8, Canada \and
   \protect\supit{\CfA}Harvard-Smithsonian Center for Astrophysics, 60 Garden St., Cambridge, MA 02138 \and
   \protect\supit{\Toronto}Dept. of Astronomy and Astrophysics, Univ. of Toronto, 50 St George St, Toronto, ON, M5S 3H4, Canada \and
   \protect\supit{\UCLA}Dept. of Physics and Astronomy, Univ. of California, Los Angeles, CA 90095
}

\maketitle

\begin{abstract}

The third-generation South Pole Telescope camera is designed to measure the cosmic microwave background across three frequency bands (95, 150 and 220~GHz) with $\sim$16,000 transition-edge sensor (TES) bolometers. Each multichroic pixel on a detector wafer has a broadband sinuous antenna that couples power to six TESs, one for each of the three observing bands and both polarization directions, via lumped element filters. Ten detector wafers populate the focal plane, which is coupled to the sky via a large-aperture optical system. Here we present the frequency band characterization with Fourier transform spectroscopy, measurements of optical time constants, beam properties, and optical and polarization efficiencies of the focal plane. The detectors have frequency bands consistent with our simulations, and have high average optical efficiency which is 86\%, 77\% and 66\% for the 95, 150 and 220~GHz detectors. The time constants of the detectors are mostly between 0.5 ms and 5 ms.  The beam is round with the correct size, and the polarization efficiency is more than 90\% for most of the bolometers.


\end{abstract}

\section{Introduction}
Measurements of the intensity and polarization of the cosmic microwave background (CMB) at high spatial resolution and high sensitivity is a powerful probe of cosmological parameters [1], inflation [2], neutrino physics [3,4], and galaxy clusters [5]. Currently there are active efforts from ground-based [6-9] and balloon-borne [10-12] projects on this front. The third-generation South Pole Telescope camera (SPT-3G)  is one of the third-generation ground-based CMB cameras designed for all these scientific goals [13]. SPT-3G uses a large-aperture wide-field optical design that fits a large focal plane with $\sim$16,000 sinuous-antenna-coupled transition-edge sensors (TESs), which gives us a high mapping speed and unprecedented sensitivity. For a broad overview of SPT-3G see [14], and for the first-year performance see [15].

 This paper is organized as follows: in section 2 we discuss the optics and pixel design for SPT-3G. We then discuss the on-site measurements of the detectors' spectral bands in section 3.1. The time constant measurements will be detailed in section 3.2, and the optical efficiency measurements will be in section 3.3. We present the detectors' beam and polarization properties in section 3.4. We summarize all the results in section 4.
\section{The optical design}
\subsection{Optical coupling}

The SPT-3G camera is coupled to the South Pole Telescope, a ten-meter off-axis Gregorian telescope located at the geographic South Pole [16]. The sky signal illuminates the primary mirror and is then reflected by a 1.8~m secondary mirror after the primary focus into a folding flat mirror (Fig.1 \textit{left}). The Gregorian focus is formed after the folding flat mirror and is right in front of the 25-inch diameter high-density polyethylene (HDPE) receiver cryostat vacuum window with triangular-groove anti-reflective (AR) coating. After the window there are multiple layers of polyethylene foam and an alumina plate at 50~K for blocking the infrared loading. Three 27-inch plano-convex lenses at 4~K re-image the Gregorian focal plane on to the detector wafers. These lenses reduce aberrations and enable a larger field of view,  compared to not having the lenses and putting the detector wafers at the Gregorian focus. The lenses are made from low-loss alumina by Coorstek and are AR-coated with laminated expanded polytetrafluoroethylene (PTFE).  Between the second lens and the third lens, where the beam waist is the narrowest, we set a Lyot stop to reduce stray reflections and put a low-pass metal mesh filter to further cut the out-of-band optical loading. Finally the sky signal is focused by the alumina AR-coated lenslets onto the individual pixel antenna. The beam size of the pixels through the Lyot stop is 30~deg, which is re-imaged by the optical system and gives us a 1.9 deg diameter field of view on-sky.
\begin{figure}[htbp]
\begin{center}
\begin{subfigure}[b]{0.22\textwidth}
                \includegraphics[height=3cm]{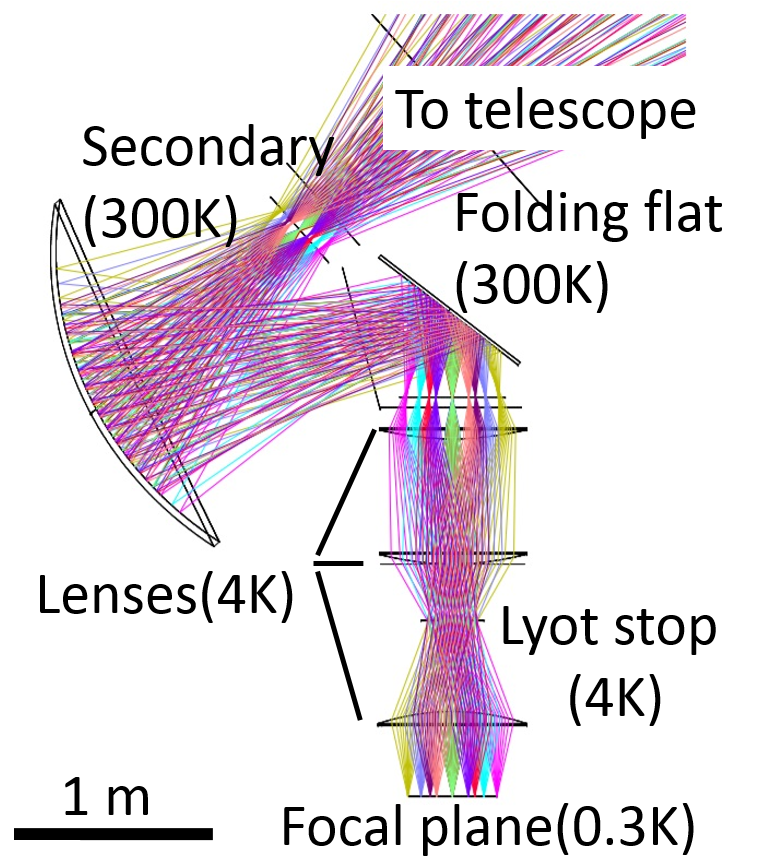}
                \label{optics}
\end{subfigure}%
\begin{subfigure}[b]{0.329\textwidth}
                \includegraphics[height=3cm]{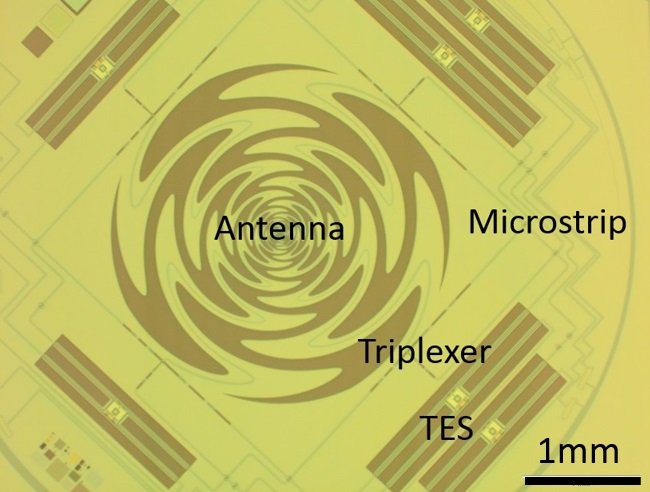}
                \label{pixel}
\end{subfigure}%
\begin{subfigure}[b]{0.40\textwidth}
                \includegraphics[height=3cm]{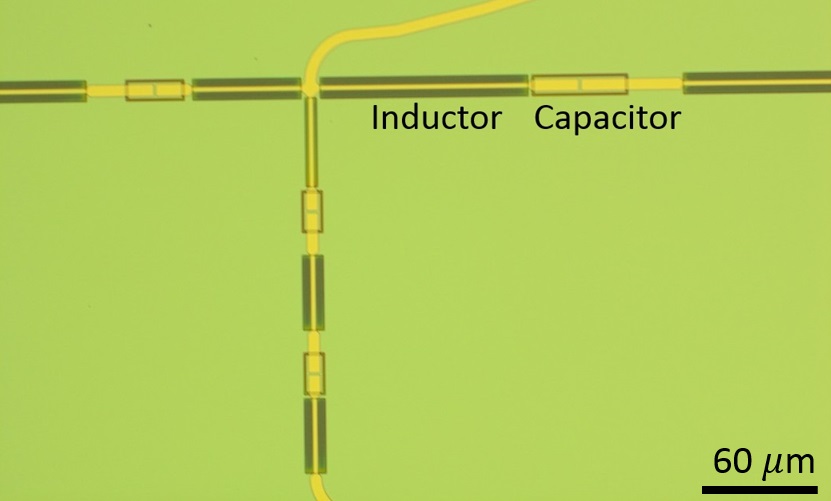}
                \label{triplexer}
\end{subfigure}%
\caption{{\it Left} is a ray-tracing diagram of the SPT-3G optical system. After the primary, a secondary ellipsoidal mirror, a flat tertiary mirror, and three lenses are used to image the sky on the focal plane. {\it Center} is a photograph of a single detector pixel. The sinuous antenna can be seen in the center. Microstrip lines and triplexers are used to connect the sinuous antenna to the six TESs. {\it Right} is a photograph of a triplexer. The inductors are coplanar waveguides with the ground plane etched, and the capacitors are two parallel plates coupled to each other. }
\end{center}
\label{fig1}
\end{figure}
\subsection{The multichroic focal plane}
The microwave radiation collected by the optical system is measured by a multichroic focal plane consisting of ten detector wafers with 269 pixels per wafer. To enhance the coupling to the detector wafer, we use alumina lenslets, one per pixel, mounted on a silicon wafer to focus the received microwave. Within each pixel (Fig. 1 \textit{middle}) in the detector wafer, a self-complementary sinuous antenna with a frequency-independent impedance over a broad range of frequency is used to couple to the microwave. Microstrip lines designed to match the antenna's driving impedance are used to transfer the collected microwave, and in-line triplexers with a trio of three-pole lumped element filters (Fig. 1 \textit{right}) are used to distribute the microwave into three different frequency bands measured by three TESs. The detailed  parameter tuning and fabrication processes  are summarized in [17-21]. The TESs are read out using a digital frequency-domain multiplexing (DfMUX) system [22-23].

\section{Optical characterization results}
\subsection{Frequency bands}

The frequency bands of the detectors were measured by a Fourier Transform Spectrometer (FTS) using a symmetric Martin-Puplett design [24]. The FTS has two input ports coupled to two beam-filling sources, one of which is a piece of Eccosorb HR10 [25] at room temperature, and the other one is a blackbody operating at 1000~K.  Within the FTS box (Fig. 2 \textit{left}), there are four pairs of mirrors for folding the beam, four polarizers for polarizing, mixing,  recombining and splitting the beam, and one moving mirror driven by a linear driver for creating an optical delay between the two optical paths. The folding design helps make the FTS's size as small as 15 by 10 by 3~inches and its weight as low as 13~lbs. The FTS output is proportional to the contrast of the two inputs, and is coupled through the three lenses in the receiver cryostat to the detectors on the focal plane. The FTS output beam is arranged to have the same f number as the detectors' beams through the lenses. The primary and the secondary mirror of the telescope are not involved in this measurement.  To avoid detector saturation, we put a Mylar beam splitter to couple only 2\% of the power from the FTS to the detectors, and redirect 98\% of the optical coupling onto the sky. To map out the focal plane more efficiently, the FTS box is put on a 2D linear driver which can move the FTS to the calculated position of the target bolometers automatically.

\begin{figure}[htbp]
\begin{center}
\begin{subfigure}[b]{0.4\textwidth}
                \includegraphics[height=3.3cm]{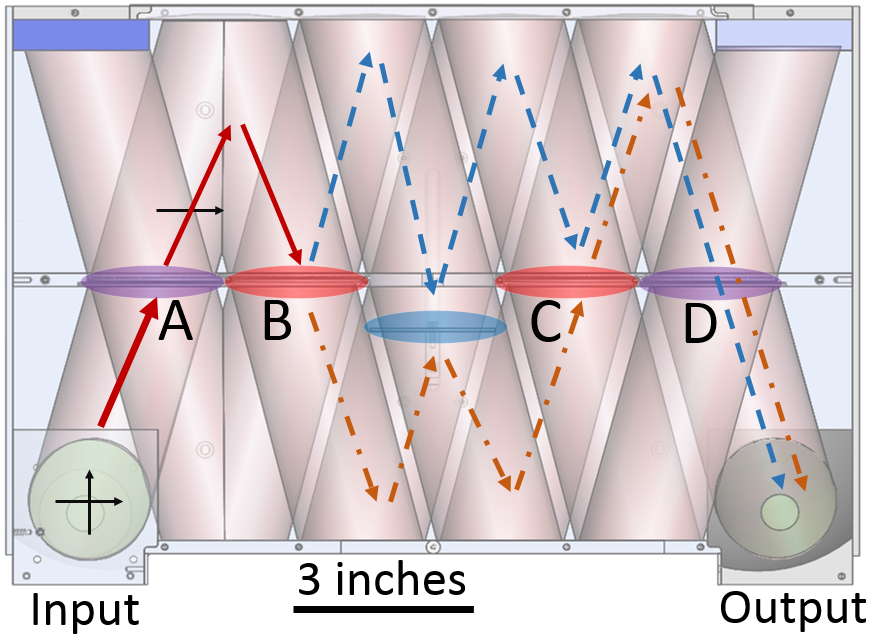}
                \label{fts_schematics}
\end{subfigure}%
\begin{subfigure}[b]{0.4\textwidth}
                \includegraphics[height=3.5cm]{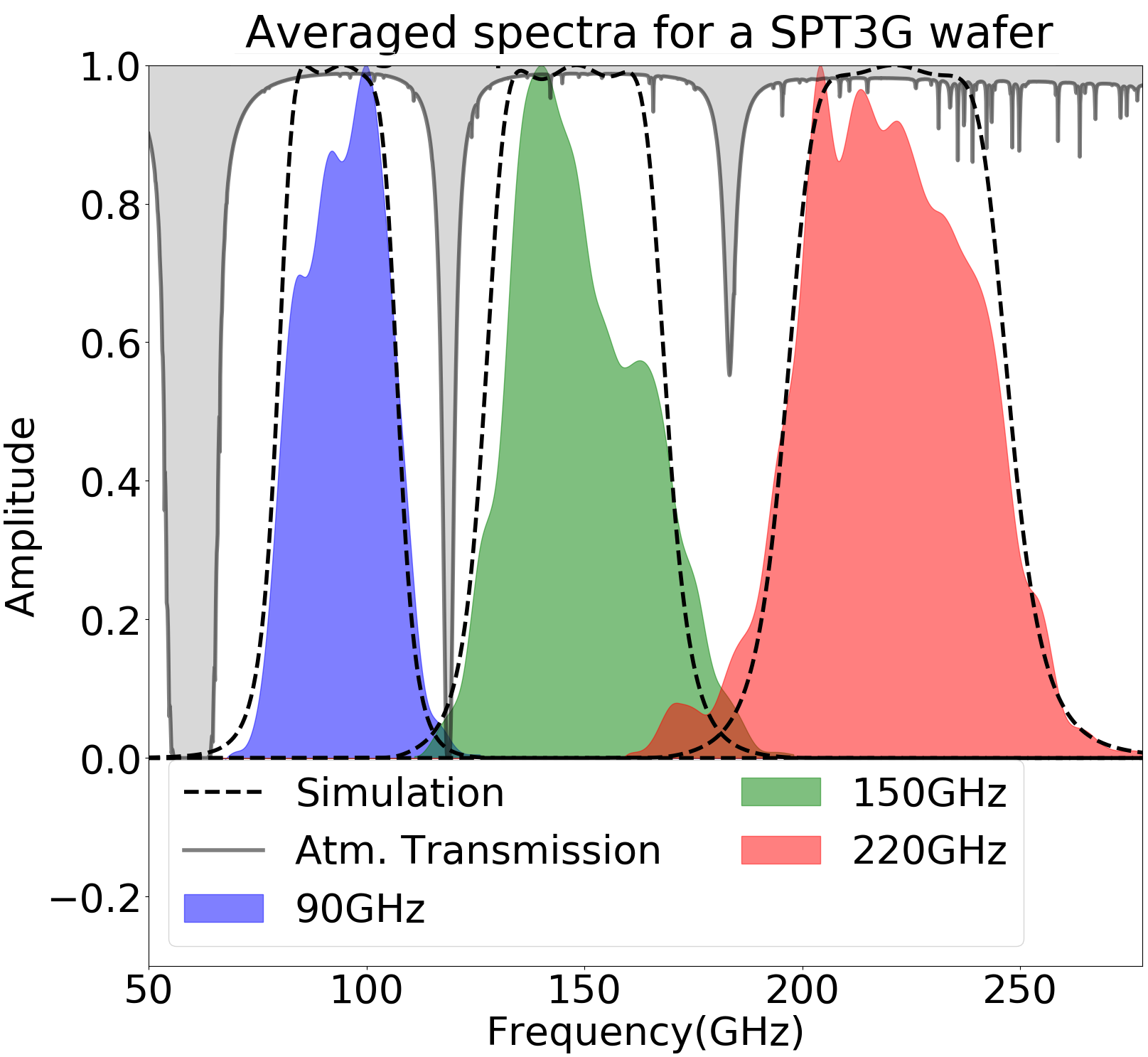}
                \label{fts_band}
\end{subfigure}%
\begin{subfigure}[b]{0.2\textwidth}
                \includegraphics[height=3.3cm]{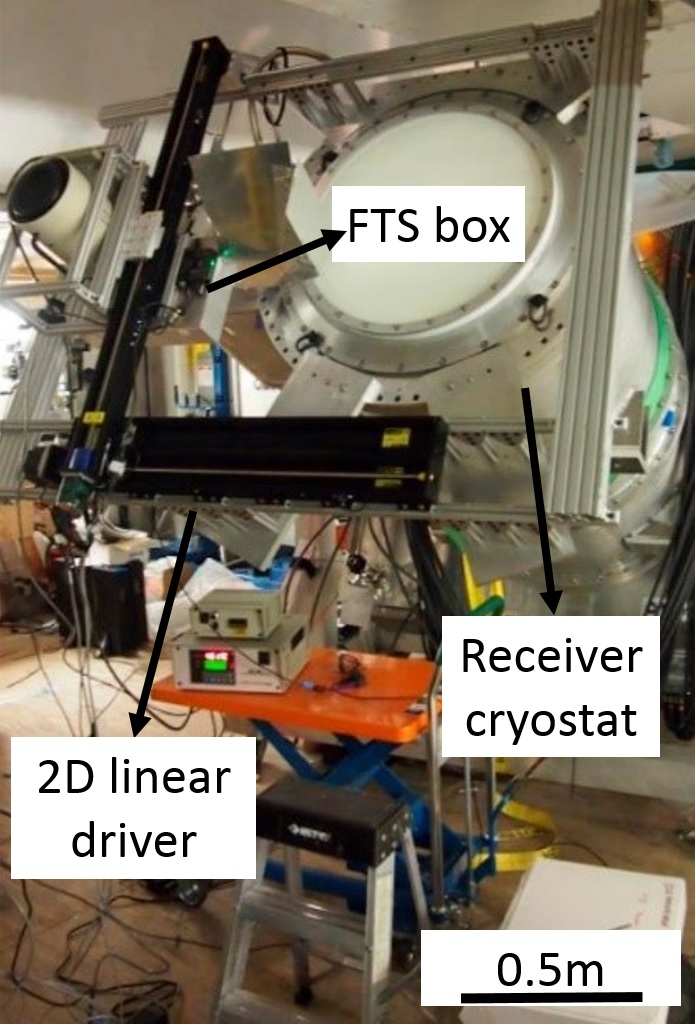}
                \label{fts_coupling}
\end{subfigure}%

\caption{{\it Left} is the schematics of the FTS. A, B, C, and D are each polarizers and act to polarize, mix, recombine and split the signal (in order of A, B, C and D). The moving mirror in the center (blue) creates the optical delay between the two optical paths. {\it Middle} is the measured averaged bands from a deployed wafer, overplotted with the simulated bands and atmospheric transmission. {\it Right} is a photograph of the testing assembly at the South Pole. The FTS is mounted on a 2D linear driver, which is attached to the receiver cryostat. }
\end{center}
\label{fig1}
\end{figure}

The recorded data is the time varying power incident on the detectors (timestream) while the FTS is scanning back and forth, and contains multiple interferograms. To extract the frequency bands from the timestream data, we first find the center of the interferograms by correlating the timestream data with a timestream template created from the guessed bands. We then separate the interferograms, do FFTs for each of the interferograms and average all the FFTs  to reduce the noise.

Fig. 2 \textit{middle} is the average frequency bands for one of the deployed detector wafers.  The bands for the other wafers are very similar. For the 95, 150, and 220~GHz bands, the measured band centers are $96\pm1$, $150\pm 3$, and $220\pm 3$~GHz, respectively, and the measured band widths are  $25\pm 2$, $33\pm 3$, and $46\pm 2$~GHz, respectively. The measured band edges agree to within 3~GHz of Sonnet simulations using the appropriate dielectric layer thickness, and the bands lie within the atmospheric transmission windows. The difference between the measured and simulated band edges is likely caused by a frequency-dependent variation of the dielectric layer \chem{SiO_2}'s dielectric constant.  The measured bands also include the transmission through the lenses, the alumina filter, and the metal mesh filter in the optical chain, which contributes to the difference between the measured and simulated band shapes.   The bands satisfy our needs.

\subsection{Time constant}

The optical time constant of the detectors describes how fast the detectors can respond to a changing optical signal. The intrinsic time constant of the detector is defined by $\tau_0= C /G$, where $C$ is the thermal capacity of the detector island, and $G$ is the thermal conductivity between the detector island and the thermal bath. However, the effective time constant is sped up to $\tau=\tau_0/(1+\mathcal L)$ due to electro-thermal feedback, where $\mathcal L$ is the loopgain of the feedback system [26]. The loopgain is dependent on the bias power and the sharpness of the TES's superconducting transition. Thus the time constant can be tuned by coupling the TES island to a deposited film with a large thermal capacity, changing the thermal link's strength or the optical loading on the TES island, and tuning the sharpness of the superconducting transition edge. The time constant needs to be larger than $5.8\tau_e \Rightarrow 5.8 ( 2   L/ R)\approx 0.35$~ms to make sure the system is stably overdamped under a perturbation [26], and smaller than $\sim 20 $~ms to be able to resolve arcminute-scale features while scanning at 1~deg/s. Here $L$ is the inductance of the multiplexing inductor and equals to \SI{60}{\,\micro\henry}. $R$ is the resistance of the TES sensor and is approximately \SI{2}{\,\ohm}.

\begin{figure}[!ht]
\floatbox[{\capbeside\thisfloatsetup{capbesideposition={right,center},capbesidewidth=4cm}}]{figure}[\FBwidth]
{\caption{{\it Top left} is a cross-section view of the calibrator model.  {\it Top right} is a histogram of the detectors' time constants.\newline \newline\newline \newline{\it Middle left} is a cross-section view of the cold load model.   {\it Middle right} is a histogram of the detectors' optical efficiency measured with the cold load.\newline \newline \newline    {\it Bottom left} is the beam measured for a 150GHz detector.    {\it Bottom right} is a polarization efficiency measurement for two 220GHz detectors within the same pixel. }\label{fig:test}}
{\includegraphics[width=3.2in]{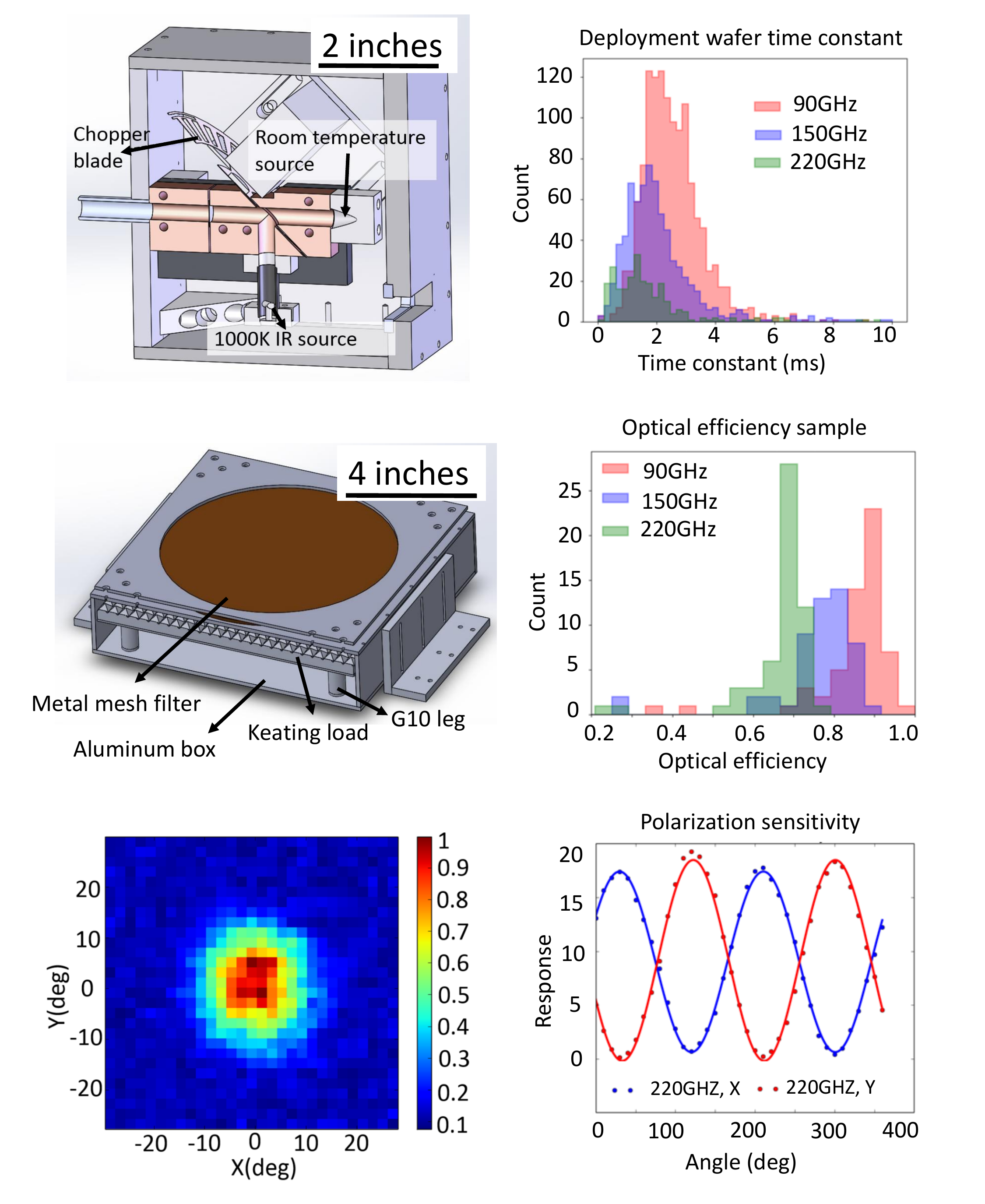}}
\end{figure}
The detectors' time constants are measured by recording the detectors' timestream data while chopping a calibrator  behind the telescope's secondary mirror. The calibrator is coupled to a hole in the center of the secondary mirror that overlaps with all detectors' beams. The calibrator has a gold-plated chopper placed at 45~deg relative to the box surface (Fig. 3 \textit{top left}). The chopper serves two functions: when the blade is in the optical path, the chopper is reflecting off the radiation from a 1000~K thermal filament; when the blade is out of the way, the chopper is transmitting the room-temperature blackbody radiation. We measure the detector response vs. chopper frequency using a phase-locked demodulator and we assume a single exponential decay model to fit the detector time constant.

According to the fit results, most detectors' time constants lie within the desired range (Fig. 3 \textit{top right}). However there are a few detectors with small time constants which are marginally stable. Reducing the detector saturation power will increase the time constants.

\subsection{Optical efficiency}

The optical efficiency of the detectors is measured in lab with a temperature-controlled cold load (Fig. 3 \textit{middle left}) placed in front of the detector wafer and is calculated by dividing the received optical power by the expected optical power assuming single-mode beam-filling coupling. The cold load used for the calibration has a piece of blackbody material [27] mounted on top of an oxygen-free high thermal conductivity copper plate suspended from the cold load box by four high-pressure fiberglass laminate (G10) legs. The temperature of the blackbody and the copper plate is controlled by a PID feedback system and ranges between 5~K and 20~K. The cold load is designed with a large aperture covered by low-pass metal-mesh filters to fill up the detectors' beam while not loading the cold stage with too much optical power. Fig. 3 \textit{middle right} shows optical efficiency measurements of a test wafer. The optical efficiency including the detector, lenslet, and lenslet AR coating, is 86\%, 77\%, and 66\% for the 95, 150 and 220~GHz detectors.

The end-to-end optical efficiency of the telescope is also measured by making sky maps of an object with known flux, like RCW38, and comparing the received power to the expected power. The measured end-to-end efficiency is 32\%, 33\%, and 9\% for the 95, 150 and 220~GHz detectors. The reduction in this measurement compared to the cold load measurement is mostly caused by the reflection and absorption of the lenses and the alumina filter. With the loss through these elements factored in, the two measurements agree with each other within 10\% for 95 and 150~GHz detectors.  The 220~GHz efficiency is a factor of $\sim$2 too low compared to expectations, and is likely caused by imperfections in the lens AR-coating.

\subsection{Beam and polarization}

The beams of the detectors need to match the beam from the telescope. We designed a beam-mapper to measure the detectors' beams. The core part of the beam-mapper is an electrically chopped thermal source [28]  mounted in the center of a metal plate coated with Eccosorb HR10. The assembly of the source and the absorptive plate is moved by two linear drivers in a 2D plane to point at the detectors from different angles and probe the beam. The FWHM beam sizes of the 95, 150 and 220~GHz detectors are measured to be 33, 20 and 14~deg, which match our expectations. These numbers are at the wafer, not at the camera input, where the beam size is set by the Lyot stop and should be independent of frequency. The polarization efficiency is measured by rotating a wire grid in front of a thermal source. Systematics from reflections are carefully reduced by placing the wire grid 45~deg relative to the optical path and coating the surroundings of the optical path with Eccosorb HR10. Most of the detectors have polarization efficiency of $>90$\%. Sample measurements are shown in Fig. 3 \textit{bottom}.

\section{Conclusion}

To summarize, we developed calibration instruments and used them to calibrate the optical performance of the SPT-3G focal plane. We made a compact Martin-Puplett FTS for measuring the detectors' frequency bands. The measured bands agree with Sonnet simulations. We designed a calibrator which can chop at different frequencies for probing the detectors' time constants, which are mostly between 0.5~ms and 5~ms. The optical efficiency was calibrated by a temperature-controlled cold load, and is 86\%, 77\%, and 66\% for the 95, 150 and 220~GHz detectors. These numbers also agree with the optical efficiency obtained by looking at RCW38. The beam characterization was done by moving a thermal source in a 2D plane in front of the detectors. The beams of the detectors have symmetric 2D Gaussian profiles and their FWHMs are 33, 20 and 14~deg for the 95, 150 and 220~GHz frequency bands. The polarization efficiency of the detectors was calibrated by a polarized thermal source and is mostly $>90$\%.

\begin{acknowledgements}
This is a pre-print of an article published in Journal of Low Temperature Physics. The final authenticated version is available online at: \url{https://doi.org/10.1007/s10909-018-1935-y} (to be published soon). The South Pole Telescope is supported by the National Science Foundation (NSF) through grant PLR-1248097. Partial support is also provided by the NSF Physics Frontier Center grant PHY-1125897 to the Kavli Institute of Cosmological Physics at the University of Chicago, and the Kavli Foundation and the Gordon and Betty Moore Foundation grant GBMF 947. Work at Argonne National Laboratory, including Laboratory Directed Research and Development support and use of the Center for Nanoscale Materials, a U.S. Department of Energy, Office of Science (DOE-OS) user facility, was supported under Contract No. DE-AC02-06CH11357. Work at Fermi National Accelerator Laboratory, a DOE-OS, HEP User Facility managed by the Fermi Research Alliance, LLC, was supported under Contract No. DE-AC02-07CH11359. NWH acknowledges support from NSF CAREER grant AST-0956135. The McGill authors acknowledge funding from the Natural Sciences and Engineering Research Council of Canada, Canadian Institute for Advanced Research, and Canada Research Chairs program.
\end{acknowledgements}

\pagebreak

\end{document}